\documentclass{emulateapj}
\usepackage{apjfonts}

\shorttitle{Ly$\alpha$ emission from GRB host galaxies}
\shortauthors{Niino, Totani \& Kobayashi}

\begin{document}

\title{Probing the Long Gamma-Ray Burst Progenitor by Lyman-alpha
Emission
of Host Galaxies}

\author{Yuu Niino, Tomonori Totani}
\affil{Department of Astronomy, School of Science, Kyoto University,
Sakyo-ku, Kyoto 606-8502, Japan}
\email{niinou@kusastro.kyoto-u.ac.jp}
\and
\author{Masakazu A. R. Kobayashi}
\affil{Optical and Infrared Astronomy Division,
National Astronomical Observatory of Japan, Mitaka, Tokyo 181-8588,
Japan}

\begin{abstract}
  Long gamma-ray bursts (GRBs) have been suggested to occur
  preferentially in low metallicity environment. We discuss the
  possibility and theoretical aspects of using Lyman $\alpha$ emission
  properties of long GRB host galaxies as a metallicity indicator of
  high redshift GRB environments, where direct metallicity
  measurements are not easy. We propose to use the fraction of
  Ly$\alpha$ emitters (LAEs) in long GRB host galaxies as a function
  of UV luminosity, which can be compared with star-formation-rate
  weighted LAE fraction of Lyman-break galaxies as the standard in the
  case of no metallicity dependence.  There are two important effects
  of metallicity dependence of long GRB rate to change the LAE
  fraction of host galaxies. One is the enhancement of intrinsic
  Ly$\alpha$ equivalent width (EW) by stronger ionizing UV luminosity
  of low metallicity stellar population, and the other is extinction
  by interstellar dust to change the observable EW.  Based on a latest
  theoretical model of LAEs that reproduce observations, we argue that
  the latter is likely to work in the opposite direction to the
  former, i.e., to decrease LAE fraction if GRBs preferentially occur
  in low-metallicity environments, because of the clumpy inter-stellar
  medium effect.  The high LAE fraction of GRB host galaxies indicated
  by observations is quantitatively be explained by the LAE model if
  GRBs occur when $Z \lesssim 0.1 Z_\odot$, although this result is
  still indicative because of the limited statistics and theoretical
  uncertainties.  This result demonstrates that the LAE statistics of
  GRB hosts may give us useful information in the future.
\end{abstract}

\keywords{gamma rays: bursts --- galaxies: high-redshift}

\section{INTRODUCTION}
Long gamma-ray bursts (GRBs; hereafter GRB means
long gamma-ray burst unless particularly mentioned)
are the brightest astronomical transient
event, giving us an important laboratory of high energy astrophysics
in extreme conditions, and an important tool to probe the high
redshift universe.  Association of some observed GRBs with
energetic type Ic supernovae (SNe) is considered to be an
observational evidence that GRBs originate from
core-collapses (CCs) of very massive stars \citep[e.g. ][]{hjo03,sta03}.
However, the occurrence rate of GRBs is much lower
than that of normal CC SNe, and the condition required for a GRB
to occur from a CC still remains as one of the most outstanding
questions about GRBs.

Theoretical studies on possible GRB progenitors and production
mechanisms of GRBs \citep[e.g. ][] {mac99,yoo05,yoo06,woo06} suggest
that low metallicity may be a necessary condition for the progenitor
to produce a GRB event. It has also been suggested from observations
that GRB host galaxies are systematically fainter than those in the
case that GRBs are unbiased tracer of star formation or CC SN rate,
indicating that GRBs may preferentially occur in low metallicity
environment because fainter or smaller galaxies generally have lower
metallicity (Le Floch et al. 2003; Fruchter et al. 2006; Wolf \&
Podsiadlowski 2007).  These interpretations have also been supported
by studies using theoretical models of galaxy formation and evolution
(Nuza et al. 2007; Lapi et al. 2008). Furthermore, Stanek et
al. (2006) reported that the distribution of metallicity of five GRB
host galaxies at $z < 0.25$ is significantly biased towards low
metallicity compared with the expectation when GRBs are unbiased star
formation tracer.

However, spectroscopic estimates of metallicity are available only for
galaxies at low redshifts ($z \lesssim 1$, Stanek et al. 2006;
Savaglio, Glazebrook \& Le Borgne 2009), while majority of GRBs occur
at higher redshift.  At higher redshifts, the metallicity of host
galaxies is discussed based on host galaxy luminosity or mass, with
the empirical relation of luminosity-metallicity and mass-metallicity
\citep[$L$-$Z$ and $M$-$Z$, ][]{fru06,che09}.  However, there is a
large scatter in the $L$-$Z$ relation.  Although the $M$-$Z$ relation
of galaxies is tighter than the $L$-$Z$ relation \citep{tre04,erb06},
rest-frame infrared luminosity must be measured to reliably estimate
galaxy stellar mass, which requires mid-infrared observations such as
those by {\it Spitzer}.  Therefore it would be useful if there is
another and independent metallicity-sensitive property of galaxies
that can relatively easily be observed even at high redshifts ($z
\gtrsim 1$) where the majority of GRBs occur. In this paper we
consider to use the Ly$\alpha$ emission of galaxies as such an
indicator.

Lyman alpha emitters (LAEs) can be detected from very large redshifts
by their strong Ly$\alpha$ emission lines \citep[][and reference
therein]{ouc08,mar08,saw08,fin09,nil09,shi09}, including the currently
highest redshift galaxy confirmed spectroscopically at $z = 6.96$
(Iye et al. 2006, Ota et al. 2008).  LAEs [typically defined as
galaxies having rest-frame equivalent width (EW) $\gtrsim$ 20 {\AA}]
are found only in metal poor galaxies in the local universe
\citep{cha93}.  Studies of spectral energy distribution (SED) of LAEs
at $z\gtrsim 3$ have suggested that they are typically younger and
less-massive than UV-continuum selected galaxies [e.g., Lyman-break
galaxies (LBGs)] at similar redshifts
\citep[e.g. ][]{gaw06,nil07,fin07,lai08}, indicating low metallicity
of LAEs.  Since low metallicity stars emit more ionizing UV photons,
it is theoretically expected that low metallicity galaxies tend to
have large Ly$\alpha$ EW. Metallicity may affect the Ly$\alpha$
properties also through extinction by interstellar dust (see \S
\ref{sec:Fesc}). Therefore, although some other effects such as
dynamics of inter-stellar medium (ISM) may well play an important role
to determine the Ly$\alpha$ properties \citep[][and references
therein]{ate08}, the probability of GRB host galaxies being LAEs could
be an indicator of metallicity dependence of GRB rate that can be used
even at high redshifts.  In fact, observations have indicated that the
LAE fraction in GRB host galaxies may significantly be higher than
that of field galaxies.  Though the sample is still small, almost all
of high redshift ($z>2$) GRB hosts are LAEs
\citep{fyn02,fyn03,jak05}, while only about 10\% -- 25\%
of LBGs at similar redshifts have such strong Ly$\alpha$ emission lines
\citep{sha03,red08}.

We will discuss the theoretical aspects of Ly$\alpha$ emission of GRB
host galaxies, and prospect for using it as a tool to get information
about GRB progenitor from future observations, especially about the
metallicity dependence, based on the latest developments of
observations and theoretical models of high redshift LAEs.  Lapi et
al. (2008) also discussed LAEs as GRB host galaxies based on the LAE
model of Mao et al. (2007), but their model assumes that `intrinsic'
ionizing photon luminosity (i.e., that originally from stars without
taking into account extinction or absorption) in each galaxy is simply
proportional to star formation rate (SFR) in the galaxy, without
dependence on age or metallicity.  It is difficult to examine the
dependence of LAE fraction on metallicity with such a modeling. Here
we use one of the latest models of LAE luminosity functions (LFs) of
Kobayashi, Totani, Nagashima (2007; 2009, hereafter KTN07 and
KTN09). This is based on a hierarchical clustering model of galaxy
formation in the framework of cosmological structure formation theory,
and the intrinsic ionizing luminosity is calculated using star
formation history taking into account metallicity evolution (see \S
\ref{sec:LyAemission}), and hence dependence of Ly$\alpha$ luminosity on
properties of stellar population can be examined in a realistic way.

In \S \ref{sec:model}, we discuss general theoretical aspects of
Ly$\alpha$ emission and its dependence on galaxy properties
and introduce the LAE model used in this work.
In \S \ref{sec:independ}, we discuss
observational strategies to extract useful information about GRB
progenitors from future data sets, with least dependence on
theoretical uncertainties.  In \S \ref{sec:depend}, we quantitatively
investigate how the fraction of LAEs in GRB host galaxies could be
affected if the GRB rate is dependent on metallicity.  In \S
\ref{sec:discussion}, we compare our model prediction to current
observed data set, and also discuss the age effect of the GRB progenitor
and the selection effect on host galaxies induced by extinction of GRB
optical afterglows.  We will summarize our conclusion in \S
\ref{sec:conclusion}.  We assume $\Lambda$CDM cosmology with
$\Omega_\Lambda=0.7$, $\Omega_M=0.3$, and $H_0=70\ \mathrm{km\ s^{-1}
\ Mpc^{-1}}$, and use the AB magnitude system throughout this paper.

\section{Theoretical Background and The Model of LAE\lowercase{s}}
\label{sec:model}

\subsection{The Galaxy Formation Model}
\label{sec:mitaka}
To construct a mock numerical catalog of galaxies, we utilize one of
the latest hierarchical clustering models (so-called semi-analytic
models) of galaxy formation by \citet{nag04}.  This model, which we
call the Mitaka model here, produces numerical catalogs of galaxies
and their evolution by semi-analytic computation of merger history of
dark matter (DM) halos and phenomenological model calculation of
evolution of baryons. The computation of DM halo merger history is
based on the standard theory of structure formation driven by cold
dark matter.  The model calculation of baryon evolution within DM
halos includes radiative cooling, star formation, SN feedback and
galaxy mergers.  Metal enrichment history (chemical evolution) of
inter-stellar medium is calculated self-consistently by the star
formation history in each galaxy.  

\begin{figure}
\includegraphics{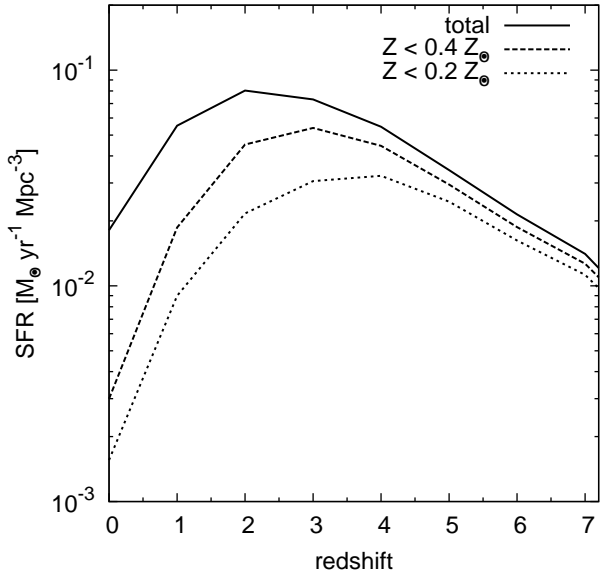}
\caption {CSFH computed in the Mitaka model. CSFH of all galaxies, and
  those of low-metallicity galaxies ($Z < 0.4 Z_{\odot}$ and $0.2
  Z_{\odot}$) are shown by solid, dashed, and dotted lines,
  respectively. 
}
\label{fig:csfh}
\end{figure}

The Mitaka model can reproduce wide variety of observed properties of
local galaxies \citep{nag04} as well as those of high redshift LBGs
\citep{kas06}.  In Fig. \ref{fig:csfh}, we show cosmic star formation
history (CSFH) computed in the Mitaka model with and without
metallicity limit.  The metallicity limited CSFHs ($Z < 0.4$ and $0.2
Z_{\odot}$) have their peak
at higher redshifts than that of total CSFH. The evolution
of cosmic SFR and dependence on the metallicity limit are similar to
those in other models utilized in studies of GRB rate evolution
\citep[e.g. ][]{dai06,lan06,lap08}, indicating that our model CSFH is
broadly consistent with the GRB flux or redshift distributions, within
the uncertainties about GRB luminosity function. 

\subsection{Ly$\alpha$ Photon Production}
\label{sec:LyAemission}
KTN07 extended this model to describe LAEs. Most of previous studies
simply assumed that Ly$\alpha$ luminosity is simply proportional to
SFR, but the KTN07 model incorporates the effect of dust extinction
with amount different from that for UV continuum, and the effect of
outflow. The KTN07 succeeded to reproduce the Ly$\alpha$ LF at various
redshifts.  Furthermore, KTN09 compared this model comprehensively
with various observed data of UV continuum LF and equivalent width
distribution of LAEs, and found a good agreement.  We use the latest
KTN09 model for the analysis in this paper, and see this paper for 
detailed description of the LAE modeling.

\begin{figure}
\includegraphics[scale=0.35]{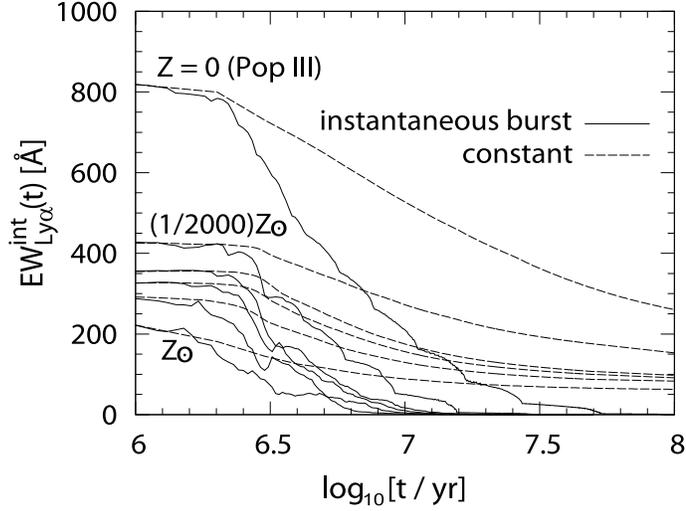}
\caption {The time evolution of intrinsic Ly$\alpha$ equivalent width
  (EW$^{\rm int}_{\rm Ly\alpha}$) of stellar populations calculated by
  the stellar population synthesis model of \citet{sch03} with the
  Salpeter IMF in the range of 1--100 $M_\odot$. The solid and dashed
  curves represent the EW$^{\rm int}_{\rm Ly\alpha}$ evolution for
  instantaneous starburst and constant star formation, respectively.
  Several curves are shown corresponding to different stellar
  metallicities: $Z = 0, (1/2000)Z_\odot, (1/50)Z_\odot,
  (1/20)Z_\odot, (1/5)Z_\odot,$ and $Z_ \odot$, from top to bottom.
  In this figure, all the ionization photons are assumed to be absorbed
  by neutral hydrogens to produce Ly$\alpha$ photons by the case B
  recombination.}
\label{fig:popsyn}
\end{figure}

There are several important ingredients to determine the Ly$\alpha$
luminosity from a galaxy. The first is the ionization luminosity
($\lambda < 912$ {\AA}) that is calculated by star formation history
and metallicity in each galaxy in the KTN09 model. Then the model
predicts the intrinsic Ly$\alpha$ luminosity assuming that a part of
ionizing photons are absorbed within the galaxy and reprocessed into
Ly$\alpha$ photons by the case B recombination.  The fraction of
ionizing photons that are absorbed by neutral hydrogen to produce
Ly$\alpha$ is treated as a part of the overall normalization factor of
Ly$\alpha$ ($f_0$, see \S \ref{sec:Fesc}).  The `intrinsic'
EW is then determined by the ratio of the intrinsic Ly$\alpha$
luminosity to the UV continuum luminosity.  Fig.  \ref{fig:popsyn}
shows the evolution of the intrinsic EW for single burst stellar
population or the case of constant star formation rate for a variety
of metallicity, calculated by the model of \citet{sch03} assuming the
Salpeter IMF in the range of 1--100 $M_\odot$.  Since stellar
populations with low metallicity emit more ionizing photons compared
with UV continuum photons around Ly$\alpha$, EW becomes larger.  This
effect is significant in young stellar populations ($\lesssim$ several
$\times 10^6$ yr). LAEs having such young age estimates have acually
been observed \citep[e.g. ][]{cha05, lai07}, and they are important to
reproduce large EW LAEs in the model of KTN09. This is one of the
important motivations to consider LAEs as the metallicity indicator of
GRB progenitors.

A caveat here is that the stellar evolution theory predicting the
ionization luminosity may still be highly uncertain at such young
ages, because of e.g., the treatment of the atmosphere of Wolf-Rayet
stars (Garc\'ia-Vargas, Bressan \& D\'iaz 1995) and stellar rotation
\citep{vaz07}. Our model is based on the widely used model of Schaerer
(2003), and qualitative trend that younger and lower metallicity
stellar population should have large intrinsic Ly$\alpha$ EW is
generally accepted.

\subsection{Ly$\alpha$ Escape Fraction and Extinction Effect}
\label{sec:Fesc}
The observable Ly$\alpha$ luminosity is determined by the intrinsic
luminosity and the escape fraction of Ly$\alpha$ photons from the
galaxy, and KTN09 considered the effects of extinction by dust.
KTN09 found that the extinction effect is
especially important to successfully reproduce various observations of
LAEs. In KTN09 model, the Ly$\alpha$ luminosity of a galaxy
including the extinction effect is computed as:
\begin{eqnarray}
L_{\rm Ly\alpha} &=& f_0 \frac{1-{\rm exp}(-\tau_{\rm Ly\alpha})} 
{\tau_{\rm Ly\alpha}}
L_{\rm Ly\alpha}^{\rm max} 
\end{eqnarray}
where $\tau_{\rm Ly\alpha}$ is an optical depth of the galaxy for
Ly$\alpha$ photons, and $L_{\rm Ly\alpha}^{\rm max}$ is Ly$\alpha$
emitted in the galaxy if all ionizing photons are reprocessed into
Ly$\alpha$ photons.  The parameter $\tau_{\rm Ly\alpha}$ is
proportional to that of UV continuum, $\tau_{\rm UV}$, and both scales
with the metal column density of galaxies calculated in the Mitaka
model.  The overall normalization factor ($f_0$) and normalization of
$\tau_{\rm Ly\alpha}$ are determined by fitting to the observed LAE
Ly$\alpha$ luminosity functions.

Because of the resonant scattering by neutral hydrogens, Ly$\alpha$
photons would take vastly different paths in a galaxy from those of UV
continuum photons, and hence $\tau_{\rm Ly\alpha}$ can be very
different from $\tau_{\rm UV}$. It is theoretically highly uncertain
whether this effect reduces or enhances Ly$\alpha$ EW, and there are
two extreme possibilities: extinction of Ly$\alpha$ photons can be
much more significant (in the case of homogeneous ISM, Ferland \&
Netzer 1979) or much less (in the case of clumpy ISM, Neufeld 1991;
Hansen \& Oh 2006) than that of UV continuum.  Recent observations by
Finkelstein et al. (2007, 2008, 2009) have indicated that the clumpy
ISM effect is in fact working in high-$z$ LAEs.  This has
theoretically been supported by KTN09 who independently found that the
clumpy ISM case (i.e. $\tau_{\rm Ly\alpha}<\tau_{\rm UV}$) 
is favored to reproduce the existence of large EW
($>240$ {\AA}) LAEs, EW-reddening correlation, and EW-$L_{\rm UV}$
correlation.  If this interpretation is correct, we expect that
galaxies with large metallicity (and hence a large amount of dust)
tend to have larger Ly$\alpha$ EW, which is in the inverse direction
to the effect of ionization luminosity mentioned above.  

It has been argued that LAEs are less dusty galaxies, both
theoretically \citep[e.g. ][]{fer79} and observationally
\citep{gaw06,gaw07,gro07,nil07,lai08,ouc08}.  These claims appear in
contradiction with the preference of the clumpy ISM dust suggested by
recent studies of Finkelstein et al. (2007, 2008, 2009) and
KTN09. However, this can be understood by Fig. 7 of KTN09 where the
plot of extinction ($A_{\rm 1500}$ at 1500 \AA) versus EW is shown for
the model and the observed data. If one selects LAEs with modest
values of EW ($\lesssim$ 100 {\AA}), the extinction is relatively
small, while the observed data of Finkelstein et al. (2009) indicate
that LAEs with large EW ($\gtrsim$ 100 {\AA}) are more dusty, which is
also reproduced by the KTN09 model. Since LAEs with EW $\lesssim$ 100
{\AA} are more abundant, a conclusion that LAEs are not dusty could be
derived depending on the sample size and selection criteria of LAEs.

\subsection{Remarks about LAE Theory in the Context of This Work}
As discussed above, the consequence of the metallicity dependence of
GRBs on the Ly$\alpha$ properties of host galaxies would be rather
complicated and theoretically still uncertain. We use the KTN09 as the
guideline in this work, but it should be noted that there are still
large uncertainties in understanding of LAEs.  Some of the theoretical
predictions may be for particular cases, or be changed by future
studies. However, our result would still be useful as the first step
to consider future possible use of LAEs to derive some information
about GRBs.  We also emphasize that the KTN09 model is one of the
latest ones, and unique to reproduce observed data consistently at
various redshifts for all of the LAE LFs in Ly$\alpha$ and UV continuum
luminosities, EW distributions, and $L_{\rm UV}$ versus EW
correlations, taking into account the selection criteria of LAEs in
each observation.

\section{Observational Strategy}
\label{sec:independ}

It is not straightforward to extract useful information from the
statistics of LAE fraction in GRB host galaxies.  Even if GRBs are
simply tracing SFR, the LAE fraction for a given sample of GRB host
galaxies is weighted by SFR of galaxies, and hence different from that
for a sample of galaxies found by flux-limited galaxy surveys.  If we
have another transient source that traces SFR, such as CC SNe,
we can compare host galaxies of GRBs and the another source.
However, LAEs are found at $z \gtrsim 2$ and CC SNe cannot be detected
at such high redshifts.

We propose to compare the LAE fraction of GRB host galaxies with the
SFR-weighted LAE fraction of high-redshift galaxies found by
flux-limited galaxy surveys. It should be possible to construct a
flux-limited sample of GRB host galaxies, with the same flux limit as
that of a galaxy survey sample.  Then we expect that the LAE fraction
of GRB host galaxies should be the same as the SFR-weighted LAE
fraction of the galaxy survey sample, provided that GRBs trace SFR. In
other words, a significant difference between the two indicates that
GRBs do not simply follow SFR in galaxies.  A caveat is possible
selection effects in detecting GRB host galaxies.  If the detection
efficiency of GRB hosts does not depend on galactic properties except
for the flux limit for galaxies, the comparison will not be
biased. However, even if the prompt and afterglow emission properties
do not depend on host properties, the detection probability of optical
afterglow could be affected by extinction in galaxies. We will discuss
this issue later in this paper (\S \ref{sec:dust}).

We need estimates of SFR of galaxies in the galaxy survey sample.  One
possibility is to use rest-frame UV luminosity as the SFR indicator,
because this is always available for galaxies found by the Lyman-break
method (i.e., LBGs). However, UV luminosity may be affected by
extinction, and a more reliable SFR estimate is favorable.  Such an
estimate is also possible, e.g., by SED fittings to multi-band
photometric data or spectroscopic observations \citep[e.g. ][]{sha05}.
The number of galaxy data sets with such rich information is
increasing thanks to the recent developments of high-redshift galaxy
searches. It should be noted that an observational SFR estimate is
generally an average of SFR until the time of the observation on a
time scale that is dependent on SFR indicators.  Here, we implicitly
assumed that the evolutionary time scale of SFR of a galaxy is longer
than the time scale of SFR indicator, so that the SFR estimate is not
biased by star formation history. This issue will be discussed in
\S \ref{sec:age}.

\begin{figure}
\includegraphics{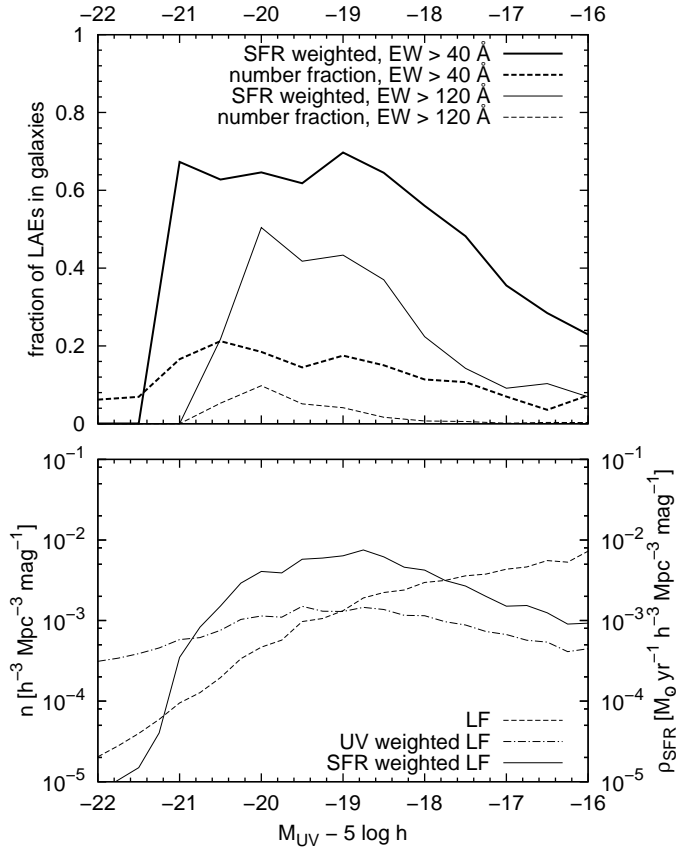}
\caption { {\it Upper Panel:} Fraction of LAEs in galaxies as a
     function of rest-frame UV ($\lambda = 1500$ \AA) magnitude,
     predicted by the KTN09 model at $z=3$. The dashed
     lines are the simple number fraction, while the solid lines are
     SFR-weighted fraction, i.e., the amount of star formation occurring
     in LAEs compared with the total star formation in galaxies.  Thick
     and thin lines correspond to two different threshold values of
     rest-frame Ly$\alpha$ EW of LAEs, 40 \AA\ and 120 \AA,  
respectively.
{\it Lower Panel}: The rest-frame UV luminosity function of
     galaxies. The dashed line is the normal LF in number, but the solid
     and dotted lines are for the SFR-weighted and UV-luminosity- 
weighted
     LFs, respectively.  The left and right ordinates are for the normal
     and SFR-weighted LF, respectively.}
\label{fig:LAEfrac}
\end{figure}

The upper panel of Fig. \ref{fig:LAEfrac} shows the simple LAE
fraction in number and SFR-weighted LAE fraction of galaxies in the
Mitaka model at $z=3$, as a function of restframe UV ($\lambda = 1500$
{\AA}) magnitude, $M_{\rm UV}$.  Two different threshold values of
Ly$\alpha$ EW are adopted as the definition of LAEs. It can be seen
that the SFR-weighted LAE fraction is significantly higher than the
simple number fraction at a fixed value of $M_{\rm UV}$. This can be
understood by the effect of clumpy dust for the extinction of
Ly$\alpha$ photons.  For a fixed $M_{\rm UV}$, galaxies with large
extinction should have larger (extinction-corrected) SFR, and large
extinction increases Ly$\alpha$ EW by the clumpy dust effect.
Therefore, even if there is no metallicity effect about GRB rates, and
GRBs simply trace SFR, we still expect a higher LAE fraction of GRB
host galaxies than the simple number fraction of LAEs in galaxy
surveys.  This result also implies that we should not use simply
$M_{\rm UV}$ as SFR indicator; if $M_{\rm UV}$ is a good SFR
indicator, we expect the same value for the SFR-weighted and simple
number fractions at a fixed $M_{\rm UV}$.

The lower panel of Fig. \ref{fig:LAEfrac} shows the UV continuum LF of
galaxies in the Mitaka model. The UV-luminosity-weighted LF and
SFR-weighted LF are also shown. The luminosity distribution of GRB
host galaxies should be the same as the SFR-weighted LF if GRBs simply
trace SFR. The SFR-weighted LF shows some deficit in the largest UV
luminosity range, compared with the UV-weighted LF. This can be
understood by the fact that galaxies with large extinction tend to
have smaller observed (i.e., extinction-uncorrected) UV luminosity,
and hence higher SFR compared with UV luminosity.  A comprehensive
comparison between the model prediction of UV continuum LF in the
Mitaka model with observations are presented in KTN09.

\section{Low Metallicity Preference of GRB\lowercase{s} and LAE
fraction}
\label{sec:depend}

Now we consider the effect of the metallicity dependence of GRBs on
LAE fraction.  As the first simple model, We assume that GRBs occur
with a rate proportional to SFR but only when the ISM metallicity of
the galaxy is smaller than a critical value, and GRBs do not occur
otherwise (the step function model):
\begin{eqnarray}
R_{\rm GRB}\propto\left\{
\begin{array}{ll}
{\rm SFR}, & \ \ Z<Z_{\rm crit} \\
0, & \ \ Z\geq Z_{\rm crit} \\
\end{array} \right..
\end{eqnarray}
This simple model of GRB rate was tested in number of studies
and reproduced various statistical
properties of GRBs such as redshift distribution and
${\rm log}N-{\rm log}P$ plot within the range of uncertainty
when it is combined with realistic model of star formation history
and metallicity of galaxies \citep[e.g. ][]{lap08}.

Our interest is in the fraction of LAEs in GRB host galaxies, and
hence we do not need to consider the normalization (the
proportionality constant) of $R_{\rm GRB}$, luminosity function
or spectrum of GRB prompt emission. Here we calculate SFR by that of
the model galaxy at the time of an observation, and implicitly assume
that the SFR evolution time scale is longer than the typical age of
the GRB progenitor.  If SFR is changing with a shorter time scale than
the progenitor age, the real GRB rate could be different from that
calculated by this formulation. This issue will be discussed in \S
\ref{sec:age}.

\begin{figure}
\includegraphics{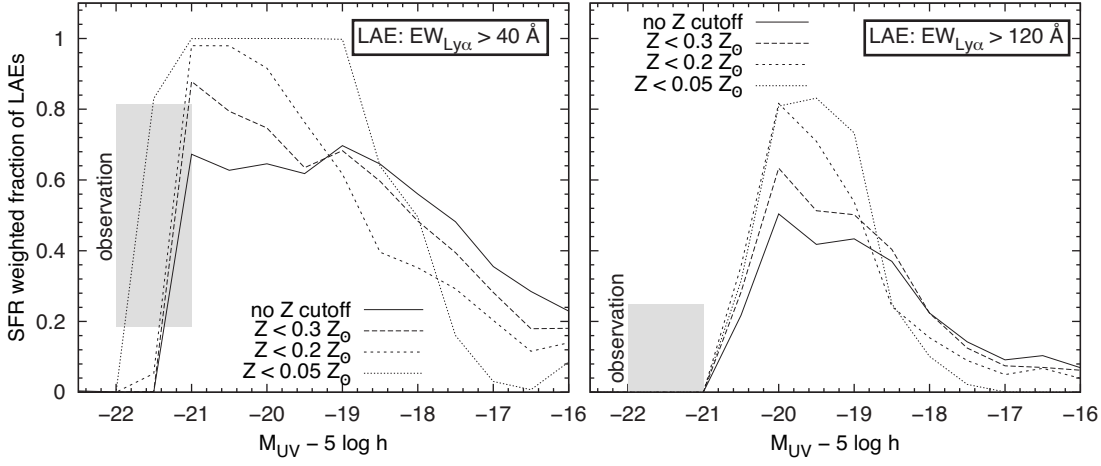}
\caption {LAE fraction in GRB host galaxies as a function of
    rest-frame UV ($\lambda = 1500$ \AA) magnitude of galaxies predicted
    by the KTN09 model at $z=3$, for various values of critical
    metallicity $Z_{\rm crit}$ (GRB occurs only when $Z < Z_{\rm crit} 
$).
    The left and right panels are for the different threshold
    values of EW$_{\rm Ly\alpha}$ for the LAE definition.  Solid line is
    the case of no metallicity dependence, which is identical to the
    solid curves of the upper panel of Fig. \ref{fig:LAEfrac}.  The 68
    \% C.L. confidence region from the observed data in Table
    \ref{tb:OBS} is shown by grey scale.  }
\label{fig:metal}
\end{figure}

We then calculate the fraction of LAEs in GRB host galaxies with
various values of $Z_{\rm crit}$ using the Mitaka model at $z=3$, and
the result is shown in Fig. \ref{fig:metal} as a function of UV
luminosity, for the two different threshold values of LAE EW as in
Fig. \ref{fig:LAEfrac}.  The LAE fraction without the metallicity
dependence ($Z_{\rm crit} = \infty$) is plotted together.  Compared
with the case of no metallicity dependence, the predicted LAE fraction
becomes higher at large UV luminosity ($M_{\rm UV} \lesssim -19$) and
lower at faint ($M_{\rm UV} \gtrsim -19$).  These trends can be
understood as follows.

Galaxies with relatively small amount of extinction are dominant in
the large UV luminosity range, since large extinction would
significantly reduce the UV luminosity.  In such a regime, the effect
of extinction by clumpy ISM on Ly$\alpha$ EWs is not significant, and
EWs are mainly determined by the stellar population, i.e., larger EWs
for younger and lower metallicity galaxies. Therefore, if GRBs trace
low metallicity stellar populations, an enhancement of Ly$\alpha$ EW
is expected. On the other hand, significantly extincted galaxies
populate the low UV luminosity range, and the effect of clumpy ISM
dust on Ly$\alpha$ EW is important.  The dust/gas ratio should be
larger for large metallicity galaxies \footnote{The amount of
  interstellar dust is assumed to be proportional to the gas
  metallicity in the Mitaka model.}, and hence our model predicts
large Ly$\alpha$ EW for metal-rich galaxies in this regime.

Therefore, the key is that UV-faint galaxies have larger extinction on
average, combined with the clumpy ISM effect.  This may sound strange
for the reader, because it is known that galaxies with larger stellar
mass tend to have larger extinction. However, we emphasize that the UV
luminosity discussed here is the apparent luminosity that is not
corrected for dust extinction.  Galaxies with larger stellar mass does
not necessarily have larger UV luminosity, due to the larger dust
extinction in UV bands.  To summarize, the two competing effects of
metallicity on the LAE fraction discussed in \S \ref{sec:model}, i.e.,
stellar population and extinction, appear in different ranges of UV
luminosity of host galaxies.

\begin{figure}
\includegraphics{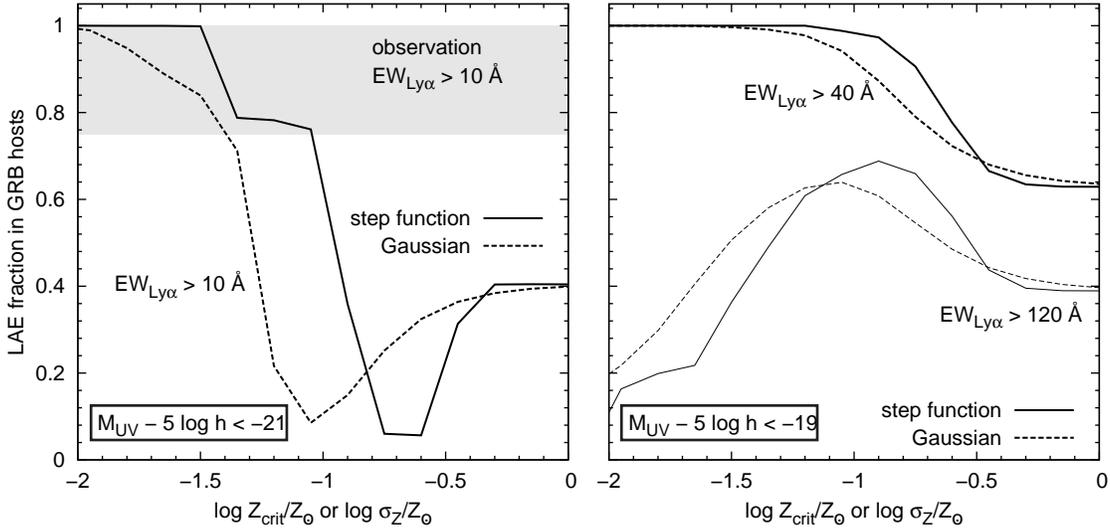}
\caption {LAE fraction in GRB host galaxies brighter than $M_{\rm
      UV} - 5 \log h = -21$ (left) and $-19$ (right) in the KTN09 model
    at $z=3$ is shown as a function of the model parameters ($Z_{\rm
      crit}$ or $\sigma_Z$) of the metallicity dependence of GRB
    rate. The solid curve assumes that GRBs occur only when $Z < Z_{\rm
      crit}$, while the dashed curve assumes a Gaussian-type metallicity
    dependence of GRB rate with the dispersion parameter $\sigma_Z$.  In
    the left panel, the 68\% confidence region by the observed data in
    Table \ref{tb:OBS} is shown by grey scale. In the right panel, two
    models with different values of the threshold EW$_{\rm Ly\alpha}$ are 
    shown by thick (40 \AA) and thin (120 \AA) lines.
}
\label{fig:vsZcrit}
\end{figure}

We also calculate the LAE fraction of GRB host galaxies integrated by
UV magnitude in the range corresponding to the observed data with the
threshold EW of 10 {\AA}, and show in the left panel of
Fig. \ref{fig:vsZcrit} as a function of the critical metallicity.  The
LAE fraction rapidly increase with decreasing $Z_{\rm crit}$ at
$Z_{\rm crit} \sim 0.1 Z_\odot$.  This metallicity scale is
interestingly similar to those suggested by theoretical studies of
stellar evolution \citep{woo06,yoo05} and spectroscopically measured
metallicity of local GRB host galaxies \citep{sta06,sav09}.  This
sensitivity of the LAE fraction for the metallicity scale of our
concern indicates an interesting opportunity to constrain the GRB
progenitor by future observations of high redshift galaxies and GRBs.

It should be noted that the behavior of LAE fraction as a function of
$Z_{\rm crit}$ is not simple, but it depends on the EW threshold and
UV magnitude range of the galaxy sample constructed.  As discussed
previously in this section, LAE fraction increases with decreasing
$Z_{\rm crit}$ in the bright UV magnitude range, and decreases in the
faint UV range.  Hence, if we consider only brightest galaxies, the
LAE fraction converges to unity due to the stellar population effect,
which is the case of the left panel of Fig. \ref{fig:vsZcrit}. (The
effect of clumpy ISM which is into the opposite direction to the
ionization luminosity effect is also seen in the range of log $Z_{\rm
  crit}\sim -0.6$ -- $-0.3$.) However, if we take fainter UV magnitude
range into account, the behavior of LAE fraction becomes complicated.
LAE fraction increases with decreasing $Z_{\rm crit}$ at large $Z_{\rm
  crit}$, but the trend turns over and it decreases with decreasing
$Z_{\rm crit}$ at small $Z_{\rm crit}$, for some values of the EW
threshold, as shown in the right panel of Fig. \ref{fig:vsZcrit}. This
point should be kept in mind when the future observations are used to
constrain the nature of the GRB progenitor.

The step function with the critical value of $Z_{\rm crit}$ is likely
to be too simple to describe the realistic GRB production efficiency
as a function of metallicity. Therefore we also try a more smoothly
varing function (the Gaussian model):
\begin{equation}
R_{\rm GRB} \propto \rm{exp}\left(-\frac{Z^2}{2\sigma_{\rm
Z}^2}\right) \times {\rm SFR},
\end{equation}
and the LAE fraction with this model is plotted against
$\sigma_{\rm Z}$ instead of $Z_{\rm crit}$
in Fig. \ref{fig:vsZcrit}.  Qualitative result is not
changed from the case of the step function model.

\section{Discussion}
\label{sec:discussion}

\subsection{Comparison to Observations}
\label{sec:observation}

Here we compare our result with observations.  Currently only one
published result is available for the LAE statistics of GRB host
galaxies (Jakobsson et al. 2005), which is based on seven GRB hosts
with Ly$\alpha$ observations including those of Fynbo et al. (2002,
2003).  Their result is summarized in Table \ref{tb:OBS}.  In these
seven host galaxies, five have strong Ly$\alpha$ emission (EW $> 10$
\AA).  Though Ly$\alpha$ emissions were not detected in two host
galaxies (000301c and 020124), the upper limits are not severe and we
cannot exclude the possibility that these two are also LAEs having EW
$> 10$ {\AA}.  It is difficult to set strong constraints on Ly$\alpha$
EW for galaxies fainter than $\sim$28 mag in optical (rest UV) bands
by the current observing facilities, and hence we do not include the
three faintest host galaxies in Table \ref{tb:OBS} (000301c, 020124
and 030323) in the following discussion.  Now we have four GRB host
galaxies at $z \sim 2$--3 with known Ly$\alpha$ EWs of 14 \AA\ -- 71
\AA.  Absolute rest UV magnitudes of these are $M_{\rm UV} \sim$ $-$22
-- $-$21. It is obvious that the statistics is not yet sufficient, 
but here we compare these results with our model
at $z=3$ as the first trial.

The observational constraints are indicated by shaded regions in Fig.
\ref{fig:metal}, using the confidence limits for small number
statistics given by Gehrels (1986). In these plots adopting EW
thresholds of 40 and 120 {\AA}, our model predictions are roughly
consistent with the observed data, giving no constraint on $Z_{\rm
  crit}$. 
However, if we take the EW threshold of 10 {\AA} and the
threshold absolute rest UV magnitude of $M_{\rm UV} - 5 \log h < -21$
to match the sample of Jakobsson et al. (2005), $Z_{\rm crit} \lesssim
0.1 Z_\odot$ seems to be favored in our model
(Fig. \ref{fig:vsZcrit}), which is interestingly consistent with other
theoretical or observational indications. However, the statistics is
obviously not sufficient, and we need to await future observations to
derive more reliable conclusions.

\subsection{Progenitor Age Effect on LAE Fraction}
\label{sec:age}

As mentioned before, we have implicitly assumed that the evolutionary
time scale of SFR in galaxies is longer than the age of the GRB
progenitor and the age of stellar population used to observationally
estimate SFR. However, if SFR is rapidly changing with a comparable or
shorter time scales than these, our argument above could be
affected. Since we expect strong Ly$\alpha$ emission from very young
stellar population having large ionizing luminosity
(Fig. \ref{fig:popsyn}), the evolutionary time scale of such galaxies
might become as short as the time scales of the GRB progenitor or SFR
indicators.

Fig. 6 of KTN09 shows the distribution of mean stellar ages of LAEs in
the Mitaka model, where the age is calculated by star formation
history of each galaxy with a weight by ionization luminosity.  Since
ionization luminosity comes from massive stars having short lifetimes,
the ionization-luminosity-weighted age is about $10^{6.6}$ yr for
stellar population having a constant SFR with a duration much longer
this time scale.  According to Fig. 6 of KTN09, LAEs with large EW
have significantly shorter age than this, indicating that their SFR is
not constant on this time scale.

If the GRB progenitors have a longer time scale to evolve into a GRB
than this scale, the age effect is not negligible.  According to the
main-sequence lifetimes of massive stars (Schaller et al. 1992;
Lejeune \& Schaerer 2001), it may be the case.  In galaxies whose mean
stellar age is significantly shorter than the age of GRB progenitor,
we expect a smaller GRB rate than simply expected from SFR because the
progenitors do not have enough time to produce GRBs, while such young
galaxies have a high probability of being LAEs.  This effect would
then lead to apparent decrease of LAE fraction of GRB host galaxies, and
hence it might be confused with the effect of the low metallicity
preference of GRBs combined with the extinction by dust in the clumpy
ISM. This is an interesting possibility, and should be kept in mind
when we analyze future data sets of GRB host galaxies, though a
quantitative prediction is beyond the scope of this work.

The typical lifetime of stars contributing to the continuum UV
luminosity at 1500--2800 {\AA}, which is a popularly used SFR
indicator, is $\sim 20$ Myr (corresponding to $\sim 10 M_\odot$ stars,
Madau, Pozzetti \& Dickinson 1998). Therefore, it is also possible
that SFR of LAEs is changing with a time scale shorter than the time
scale of the SFR indicator. One must be careful about this when the
SFR-weighted LAE fraction of field galaxies is calculated as the
reference value corresponding to the case of the SFR-tracing GRB rate.

\subsection{Dust Extinction of Optical Afterglows}
\label{sec:dust}

In most cases, detection of an optical afterglow is required to
identify the host galaxy of a GRB at high redshift, and hence a sample
of identified GRB host galaxies may be biased against dusty host
galaxies in which afterglow light is significantly attenuated.  Since
the extinction should have a significant effect on the Ly$\alpha$ EW
of host galaxies as discussed in \S \ref{sec:model}, the extinction
bias of GRB host galaxies could affect the LAE statistics of them.

\begin{figure}
\includegraphics{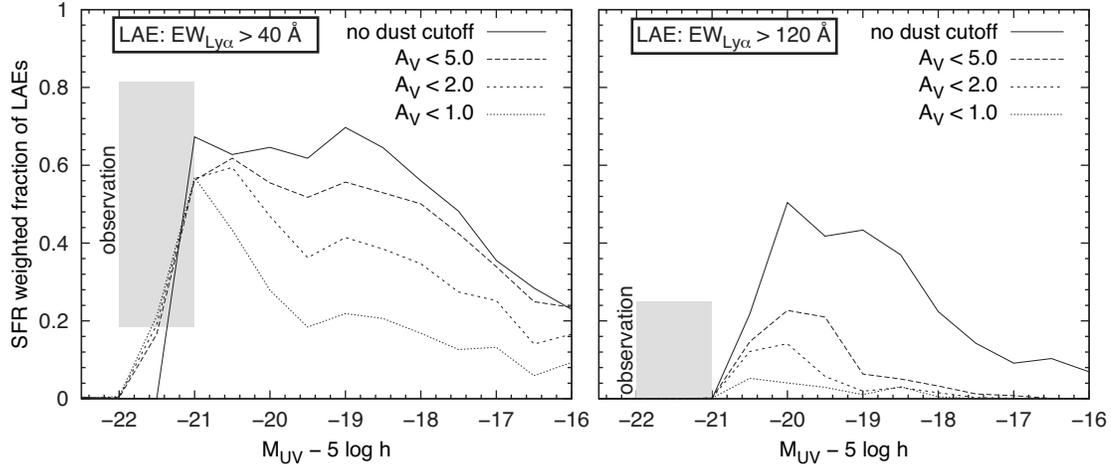}
\caption {
     The same as Fig. \ref{fig:metal}, but here GRB afterglows
     are assumed to be detected only when $A_V$ of their host galaxies  
is
     smaller than a critical value.  The GRB rate is assumed to simply
     trace SFR without metallicity dependence.}
\label{fig:dust}
\end{figure}

To investigate this effect, we employ a simple method similar to that
used to investigate the low metallicity preference of GRB events.  We
assume that we can identify the host galaxy of a GRB only when the
extinction of the galaxy is less than a critical value,
and hence the effective GRB rate becomes:
\begin{eqnarray}
R_{\rm GRB}\propto\left\{
\begin{array}{ll}
     {\rm SFR}, & \ \ A_{V}<A_{V, \rm crit} \\
     0, & \ \ A_{V}\geq A_{V, \rm crit} \\
\end{array} \right..
\end{eqnarray}
The LAE fraction of GRB host galaxies at $z=3$ predicted by this model
is shown in Fig. \ref{fig:dust}, and the LAE fraction decreases as the
afterglow extinction effect becomes strong (smaller $A_{V, \rm
    crit}$). This is simply a manifestation of the effect of extinction
by clumpy ISM dust on LAEs, i.e., enhancement of LAE EW by extinction,
which is assumed in the model of KTN09.  The change of the LAE
fraction is into the same direction in all the magnitude range, and it
can in principle be distinguishable from the effect of low metallicity
preference of GRBs showing increase and decrease of the LAE fraction
with decreasing $Z_{\rm crit}$ at $M_{\rm UV} - 5 \log h \lesssim -19$
and $\gtrsim -19$, respectively (Fig. \ref{fig:metal}).

\citet{fyn03} pointed out that the bias against dusty galaxies by the
extinction of optical afterglows may be the reason of the observed
high LAE fraction of GRB host galaxies, because it has often been
argued that LAEs are less dusty galaxies. However, as discussed in \S
\ref{sec:model}, recent studies have suggested that the effect of
extinction on LAE EW is rather inverse and to increase EW by
extinction in clumpy ISM.  Therefore, if the indication of the clumpy
ISM extinction is correct, the extinction of GRB afterglows is
unlikely to cause large LAE fraction among host galaxies.

\section{CONCLUSIONS}
\label{sec:conclusion}

We discussed the theoretical aspects of using Ly$\alpha$ emission
properties of GRB host galaxies to get implications for the GRB
progenitor. The fraction of LAEs in a sample of GRB host galaxies can
be compared with the SFR-weighted LAE fraction of a sample of LBGs at
similar redshift with a similar magnitude limit, and a significant
difference, if observed, would indicate that GRBs do not simply
follow SFR and a new physical parameter, such as metallicity, is
hidden in the relation between GRB rate and SFR.

To make quantitative predictions we used one of the latest theoretical
model of LAEs in the framework of hierarchical galaxy formation, which
is successful to reproduce a variety of recent observed data about LAE
statistics. We then tested the case that GRBs have preference to low
metallicity environments. We found that there are two effects of the
metallicity dependence on the LAE fraction of host galaxies.  One is
the intrinsic EW enhancement by low metallicity; low metallicity
stellar populations have larger ionizing UV luminosity and hence
larger Ly$\alpha$ EWs. The other is the extinction by ISM dust; low
metallicity galaxies are expected to be less dusty and dust affects
LAE EWs. According to the KTN09 model as well as recent observations,
Ly$\alpha$ EWs may be enhanced because of extinction by dust in clumpy
ISM. Therefore, the low metallicity preference of GRBs would decrease
the LAE fraction by this effect, which is into the opposite direction
to that of the intrinsic EW enhancement. We predicts that the
intrinsic EW enhancement is more significant for galaxies with large
UV luminosities, while the extinction effect dominates in UV-faint
galaxies, separated by the border of $M_{\rm UV} - 5 \log h \sim -19$
at $z=3$.

The observational indication of higher LAE fraction for GRB host
galaxies reported by Fynbo et al. (2002, 2003) and Jakobsson et
al. (2005) can be explained by our model by the intrinsic
EW enhancement effect, if GRBs occur only in low metallicity
environment of $Z \lesssim Z_{\rm crit} = 0.1 Z_\odot$. However, the
statistics is still poor and further systematic observations of GRB
host galaxies are highly desired.

We have also discussed the selection effect of GRB host galaxies by
extinction of GRB afterglow flux, which is another possible
explanation of large LAE fraction. However, recent observations and
theoretical modeling indicate that the effect of extinction is to
enhance Ly$\alpha$ EW by clumpy ISM, and in this case the selection
effect will decrease the LAE fraction.

It should be noted that the theory of LAEs is still highly uncertain,
and the quantitative predictions by the KTN09 model suffers from such
uncertainties. However, the quantitative behaviors about LAE emission
of GRB hosts presented here based on one of the latest models of LAEs
would be useful to start consideration about using LAEs to derive
information about GRBs by the future data set. Observations of high
redshift LAEs are rapidly developing, and near-future observations of
LAEs will provide us with further knowledge about this galaxy
population.  We tried to outline the important theoretical effects and
aspects that should be considered when one uses Ly$\alpha$ properties
of GRB host galaxies to get any implications about GRBs from future
observations.

\acknowledgements
This work was supported by the Grant-in-Aid for the
Global COE Program "The Next Generation of Physics, Spun from
Universality and Emergence" from the Ministry of Education, Culture,
Sports, Science and Technology (MEXT) of Japan.  TT was also supported
by the Grant-in-Aid for Scientific Research (19047003, 19740099) from
MEXT.  MARK was supported by the Research Fellowship for Young
Scientists from the Japan Society for the Promotion of Science (JSPS).
The numerical calculations were in part carried out on SGI Altix3700
BX2 at Yukawa Institutefor Theoretical Physics of Kyoto University.

%\clearpage

\begin{table}
\begin{center}
\caption{long GRB host galaxies with Ly$\alpha$ EW measurements
\label{tb:OBS}}
\begin{tabular}{llllll}
\tableline\tableline
GRB & redshift & host mag
& $F_{\rm Ly\alpha} (\mathrm{erg\
s^{-1}\ cm^{-2}})$ &
$L_{\rm Ly\alpha} (\mathrm{erg\ s^{-1}})$ & Ly$\alpha$ EW$_{\rm rest}$
(\AA)\\
\tableline
971214 & 3.42 & $R = 26.5$ & $\sim 4.5\times 10^{-17}$ & $\sim
5.0\times 10^{42}$ & $\sim 14$\\
000301c & 2.04 & $R = 28.0$ & $< 3.5\times 10^{-17}$ & $< 1.1\times
10^{42}$ & $\lesssim 150$\\
000926 & 2.04 & $U = 24.9$ & $1.6\times 10^{-16}$ & $5.1\times
10^{42}$ & $71$\\
011211 & 2.14 & $R = 24.9$ & $2.8\times 10^{-17}$ & $1.0\times
10^{42}$ & $21$\\
020124 & 3.20 & $V = 29.3$ & $< 1.5\times 10^{-17}$ & $< 1.4\times
10^{42}$ & $\lesssim 22$\\
021004 & 2.34 & $B = 24.4$ & $2.5\times 10^{-16}$ & $1.1\times
10^{43}$ & $68$\\
030323 & 2.14 & $V = 28.0$ & $1.2\times 10^{-17}$ & $\sim 1.3\times
10^{42}$ & $\sim 145$\\
\tableline
\end{tabular}
\tablecomments
{Host magnitude of GRB 020124 is from \citet{blo02}, and other values
are
from \citet[][see their Table 4 and references therein]{jak05}.
The magnitudes have been converted to AB magnitudes.}
\end{center}
\end{table}

%\clearpage

%\clearpage

%\clearpage

%\clearpage

%\clearpage

%\clearpage

\end{document}